\documentclass{cimento}
\usepackage{graphicx}
\issue{115}{070809}{July-September 2000}
\shorttitle{Electromagnetic fields around black holes and Meissner
effect} \shortauthor{Ji\v r\' \i\ Bi\v c\' ak \atque Tom\'a\v{s}
Ledvinka} 
\title{Electromagnetic fields around black holes and Meissner effect}
\author{Ji\v r\' \i\ Bi\v c\' ak \atque Tom\'a\v{s} Ledvinka}
\instlist{\inst{} Institute of Theoretical Physics\\
Faculty of Mathematics and Physics\\
Charles University\\
V Hole\v{s}ovi\v{c}k\'ach 2, 180 00 Prague 8\\
Czech Republic}
\PACSes{\PACSit{04.70.-s}{Physics of black holes}
        \PACSit{97.60.Lf}{Black holes}
        \PACSit{04.40.Nr}{Einstein-Maxwell spacetimes, spacetimes with fluids, radiation or classical fields}

}
\begin{document}

\maketitle

\begin{abstract}
The work on black holes immersed in external stationary magnetic
fields is reviewed in both test-field approximation and within
exact solutions. In particular we pay attention to the effect of
the expulsion of the flux of external fields across charged and
rotating black holes which are approaching extremal states.
Recently this effect has been shown to occur for black hole
solutions in string theory and Kaluza-Klein theory.
\end{abstract}

\section{Introduction}

Although much insight into the structure of external
electromagnetic fields around black holes was obtained in 1970s
and 1980s already, an interest in this theme continues being supported by the
appearance of new motivations. To those coming from classical
relativity belongs the ``membrane paradigm'' \cite{Th} -- in the
following (Section 4) we shall briefly discuss its validity for almost
extreme black holes. Another ``classical'' issue is the behaviour
of fields on the Cauchy (inner) horizons of charged or rotating
holes \cite{BO}. Astrophysically, the discovery of microquasars in
our Galaxy \cite{MR} makes the mechanism of the energy extraction
from rotating black holes testable more directly than it has been
possible with supermassive black holes in distant galactic
nuclei. We refer to the contributions in these proceedings
(and literature quoted therein) by M. Camenzind and R. Khanna and
by B. Punsley (see also \cite{Pu}) on the role of -- still not
properly understood -- Blandford-Znajek and related mechanisms of electromagnetic
extraction of rotational energy; Hyun Kyu Lee considers such
processes to explain gamma ray bursts (see also \cite{HK}).
In Section 5 exact spacetimes with black holes in strong fields will be
considered (we shall also mention the possibility of the existence of a
``cosmic supercollider'' formed by a supermassive black hole
surrounded by a superstrong magnetic field \cite{K}).

Finally, new motivations have appeared with studies of black hole solutions in
spacetimes with the dimensions either lower or higher than four
within the framework of various ``model'' or ``unified theories''. A
useful survey of these ``generalized'' black holes, in fact of
essentially all important developments in physics of black holes
up to 1998 is contained in the new monograph by V. Frolov and I.
Novikov \cite{FN}. In Section 6 we shall outline some new results on the expulsion
of magnetic (or more general) gauge fields from extremal black
holes in string theory.

Before mentioning these new developments we shall briefly review
our earlier results on coupled perturbations of
Reissner-Nordstr\"{o}m black holes (Section 2) and on the fluxes of
magnetic fields across black holes (Section 3).

\section{Perturbations of charged black holes}

The study of the perturbations of charged, non-rotating black
holes can be motivated from several viewpoints: (i) the
Reissner-Nordstr\"{o}m solution is the simplest solution
involving a limit - when the charge $Q$ and mass $M$ satisfy the
condition of extremality $(Q^2= M^2)$ - beyond which the horizon
is absent; (ii) the holes' interiors contain the Cauchy horizons
where generic perturbations should imply instability; (iii) the
electromagnetic perturbations are in general coupled with the
gravitational perturbations. This leads to such intriguing
effects as the conversion of gravitational waves into
electromagnetic ones, or the appearance of closed magnetic field
lines caused by ``effective gravitational currents''; the
coupling can also be used to study rigorously the motion of a black
hole under an external, non-gravitational influence. Some of these
effects will be discussed in the following.

The necessity of the coupling can easily be understood from the
perturbed Einstein-Maxwell equations written symbolically in the
form

\begin{equation}
\delta G = 8\pi \delta T,
\label{1}
\end{equation}
where the perturbed energy-momentum tensor, $\delta T \sim F^{(0)}
\delta F$, is {\it linear} in perturbations because of the
presence of the background Coulomb-type electric field $F^{(0)} \sim
Q/r^2$. Hence, the left-hand side of (\ref{1}), the perturbed
Einstein tensor, is linear in metric perturbations $\delta g$ (and
their derivatives). The perturbed Maxwell equations, $F_{\alpha
\sigma}^{;\sigma}= 4\pi j_{\alpha}$, turn out to read explicitly
as follows \cite{BD}:

\begin{equation}
(\delta F_{\alpha \sigma })^{; \sigma} = 4\pi \delta j_{\alpha} +
(\delta j_{\alpha})_{grav.},
\label{2}
\end{equation}
where the effective ``gravitational current'' is given by

\begin{equation}
\left(\delta j_{\alpha}\right) _{grav.} = F^{(0)}_{\alpha \rho; \sigma}
h^{\rho \sigma} + F^{(0)\rho \sigma} h_{\alpha \rho ; \sigma}+
{{F^{(0)}}_{\alpha}}^{\rho} \left( h^{\sigma}_{\rho;\sigma}  -\frac{1}{2}h^{\sigma}_{\sigma; \rho} \right),
\label{3}
\end{equation}
where $h_{\rho \sigma} \equiv \delta g_{\rho \sigma}$. Since in
perturbed Einstein's equations (\ref{1}) and Maxwell's equations
(\ref{2}), (\ref{3}) the electromagnetic and gravitational
perturbations are coupled, it was not an easy task to convert the
formalism into a tractable form. We shall now very briefly sketch
the history of this issue, referring to the monographs \cite{Ch},
\cite{FN} and to our extensive work \cite{Bi} and review
\cite{B} for details and citations of the original papers. A few recent
references will be given below.

The basic theory of interacting perturbations of the
Reissner-Nordstr\"{o}m black hole was started by Zerilli in
1974 who extended the Regge-Wheeler-Vishveshwara-Zerilli theory
of perturbations of the Schwarzschild black hole. After
decomposing perturbations into the (vector and tensor) harmonics,
Zerilli chose a coordinate system according to the Regge-Wheeler
gauge conditions. In this gauge he reduced the equations for
perturbations for each multipole with $l \geq 2$ to two equations
of the second order for two functions, the knowledge of which is
sufficient to determine all perturbations. Sibgatullin and
Alekseev, using a different gauge, found a pair of decoupled wave
equations in case of both parities. A novel approach to
investigate coupled perturbations was developed by Moncrief who,
by employing the Hamiltonian formalism, was able to find gauge
independent canonical variables in terms of which all metric and
electromagnetic perturbations can be determined after a gauge is
specified. The possibility to fix the gauge only towards the end
of calculations is advantageous not only from a principle,
``theoretical'' point of view. In \cite{BiP} we were able to find
explicitly all perturbations in the problem of the motion of a
charged black hole in an asymptotically uniform weak electric
field only because we chose a gauge different from the
Regge-Wheeler gauge. This choice was done after the equations for
gauge-independent quantities were solved. Moncrief also indicated
how the pair of decoupled wave equations can be obtained for
suitable combinations of gauge independent variables for
multipoles with $l \geq 2$.

In our extensive paper \cite{Bi} we developed the formalisms for the
interacting perturbations of the Reissner-Nordstr\"om black holes
in detail and clarified the relations between them. First we
expanded Moncrief's theory: all Hamilton equations following
from four Moncrief's Hamiltonians (for $l \geq 2,~ l = 1$, and
in both parities) are derived in suitable forms and from them the
wave equations for gauge invariant perturbations are obtained.
Starting then from the Hamilton equations and employing the
relations between the standard form of perturbations, $\delta
g_{\mu \nu}$ and $\delta F_{\mu \nu}$, and canonical variables,
one can express all $\delta g_{\mu \nu}$'s and $\delta F_{\mu
\nu}$'s in terms of gauge invariant variables (satisfying the
wave equations) in a suitable gauge. In \cite{Bi} this is done in
Regge-Wheeler gauge for $l\geq 2$ perturbations and in another
suitable gauge for the dipole perturbations. These results
enabled us to treat various perturbation problems (as, for
example, the expulsion of the field lines illustrated in Figure
1) and also to establish a detailed relation between the standard
(Zerilli-type) perturbation formalism and the canonical
(Moncrief-type) treatment of perturbations.

Of course, the coupled perturbations of the Reissner-Nordstr\"om
black holes were analyzed also in the framework of the
Newman-Penrose formalism which proved to be so efficient in the
case of rotating black holes. Lun, Chitre, Lee and Chandrasekhar
(see \cite{Ch}, \cite{Bi} or \cite{B} for references) are among
the main contributors to the theory. Their results are extended
in \cite{Bi} in the following points: (i) the fundamental
perturbation variables which satisfy decoupled equations are not
only coordinate gauge invariant but also invariant under
infinitesimal transformations of the Newman-Penrose tetrads; (ii) the
dipole perturbations are analyzed in the Newman-Penrose formalism
for the first time, and they are treated simultaneously with the
$l\geq 2$ perturbations; (iii) the relations between the
canonical and the Newman-Penrose basic quantities are
established.

We are recalling these results not only because they form the
theoretical framework for solving the problems like the motion of
a charged black hole in an asymptotically uniform electric field
\cite{BiP}, or the fields of stationary sources on the
Reissner-Nordstr\"om background \cite{BD}. A somewhat
``central-European'' character of the journal in which \cite{Bi}
appeared, brings its ``fruits": in 1995 the formalism for the dipole
odd-parity perturbations of the Reissner-Nordstr\"om solution was
redeveloped in \cite{B1} and in 1999 the even parity case was treated
\cite{B2} without a realization that this was done in \cite{Bi}
within the framework of three different formalisms. We
believe that in \cite{Bi} the most complete discussion is given
of the interconnections between the standard, the canonical and
the Newman-Penrose formalisms even for perturbations of
Schwarzschild black holes.

Now there exist simple exact stationary multipole solutions for coupled
perturbations of the Reissner-Nordstr\"om black holes \cite{BL}.
(Some of these solutions have very recently  been employed to study
the electromagnetic Thirring effects \cite{PF}.) In \cite{BD} we
used these solutions to construct the magnetic field of a current
loop (magnetic dipole) placed axisymmetrically on the polar axis
of the extreme Reissner-Nordstr\"om black hole. The
electromagnetic and gravitational field occurring when the
general Reissner-Nordstr\"om black hole is placed in an
asymptotically uniform magnetic field was also derived.

\begin{figure}[h]
\centering
\includegraphics[width=.49\textwidth]{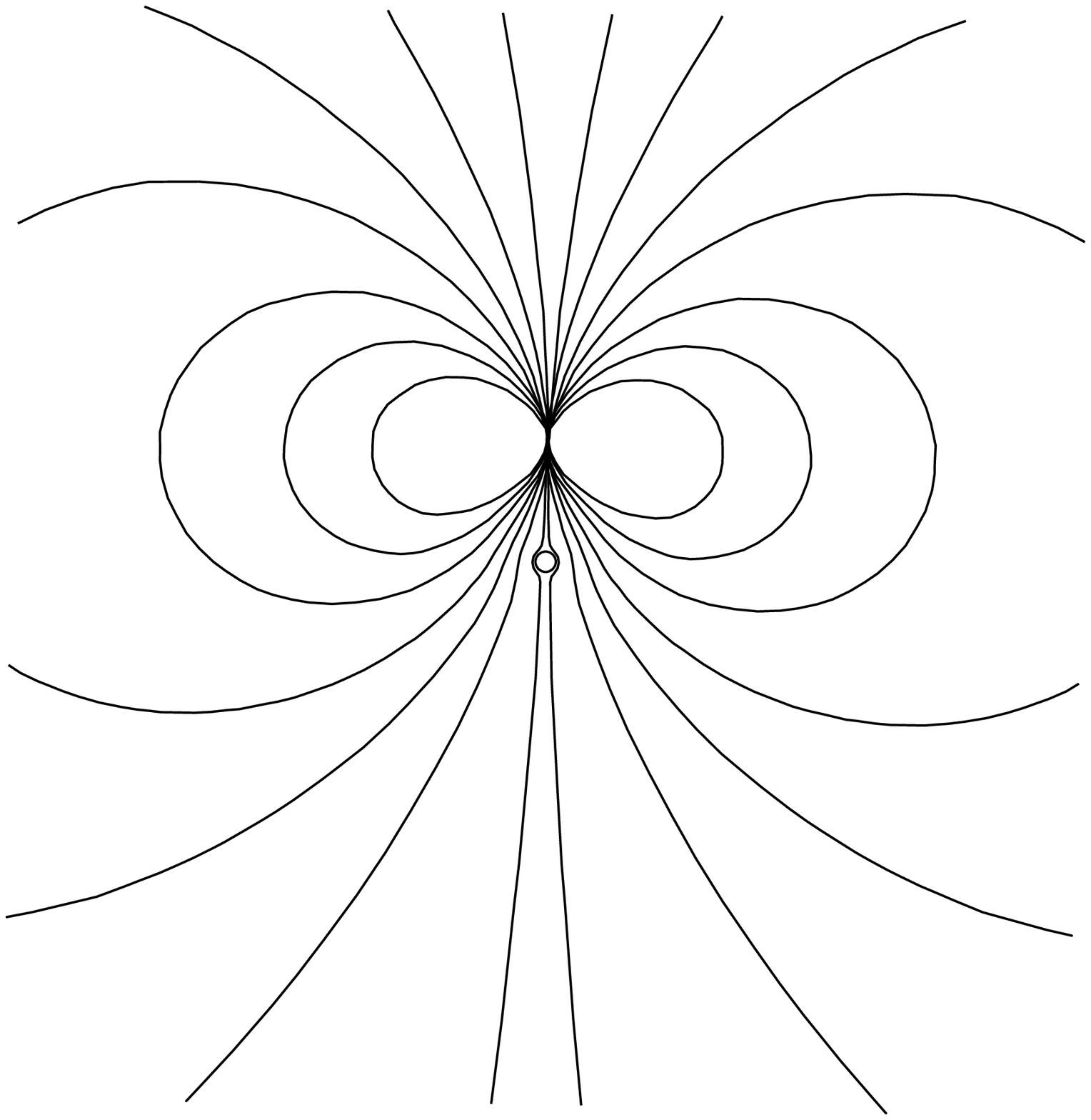}
\includegraphics[width=.49\textwidth]{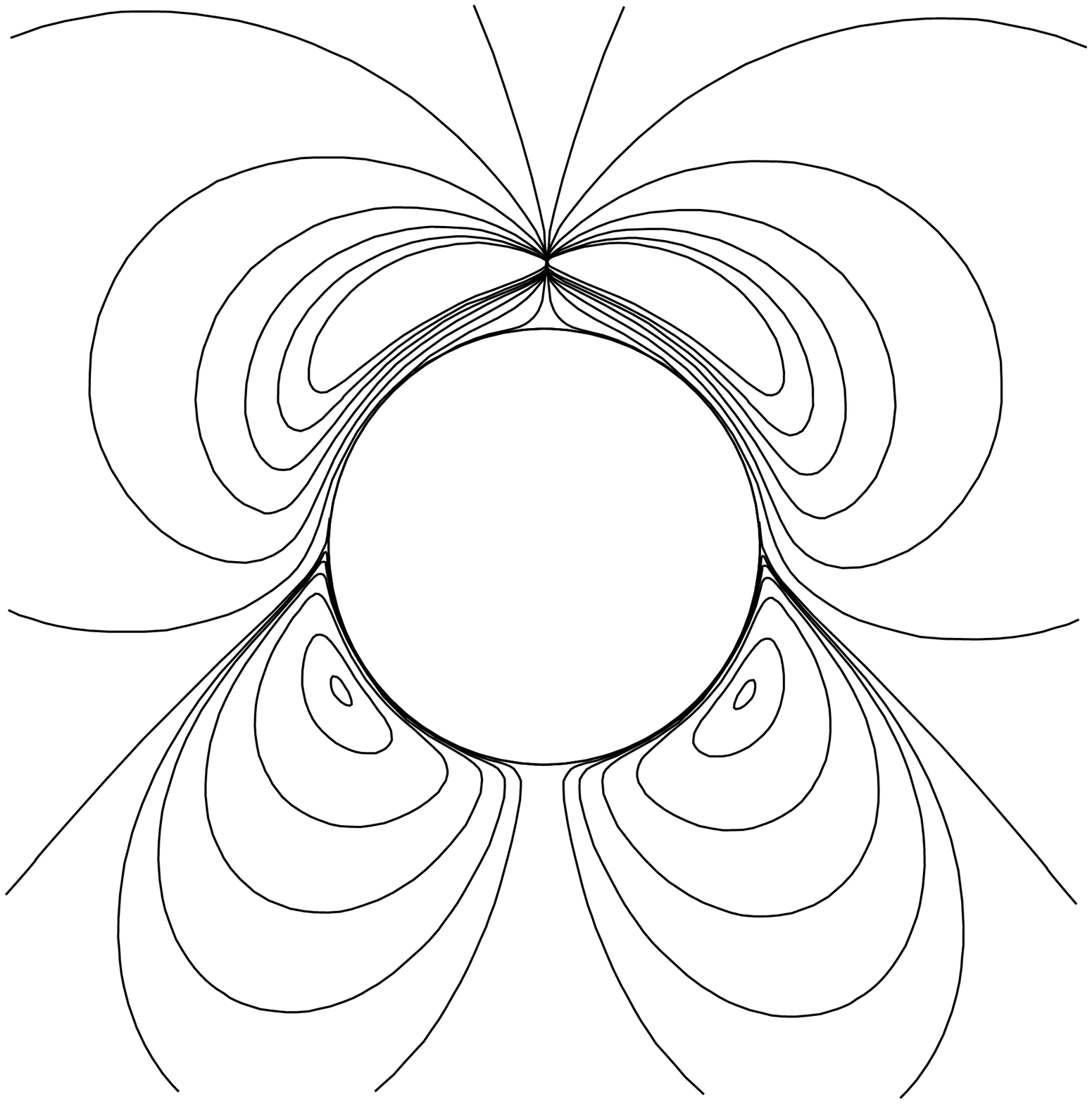}
\caption{
Field lines of the magnetic dipole placed (a) far away from
the extreme Reissner-Nordstr\"{o}m black hole, (b) close to the hole (from \cite{BD}).
}
\label{Fig1}
\end{figure}

The magnetic lines of force, as introduced by Christodoulou and Ruffini
(see \cite{CR} and references therein), were constructed
numerically for all the sources mentioned above at various
positions. We refer to \cite{BD} for details. Here, as an
illustration, we present Figure 1, showing magnetic lines of
force of the small current loop (magnetic dipole) located
axisymmetrically on the polar axis of the extreme
Reissner-Nordstr\"om black hole. One can make sure that
the structure of the closed lines of force in the region ``opposite'' to
the place where the magnetic dipole is located (Figure \ref{Fig1}b), is caused by the
coupling of electromagnetic and gravitational perturbations.
Owing to the extreme character of the black hole, no line of
force crosses the horizon.

\section{The flux of stationary magnetic fields across rotating black holes}
In order to investigate the structure of an asymptotically
uniform test magnetic field on the background of a Kerr black hole we
can start from the field $F_{\mu \nu}$ given explicitly in
\cite{BJ}. Without repeating here complicated formulas, let us
recall that each $F_{\mu \nu}$ can be expressed as a sum of two
terms, one being proportional to $B_0$, the magnitude of the
component of the field asymptotically aligned with the hole's
rotation axis, the other, $B_1$, being the magnitude of the
component  perpendicular to the axis. Following Christodoulou and
Ruffini \cite{CR} we define the magnetic (electric) lines of
force as the lines tangent to the direction of the Lorentz force
experienced by a test magnetic (electric) charge at rest with
respect to the locally non-rotating frame. For the magnetic field
lines this definition yields $dr/d\theta = -F_{\theta \phi}/F_{r
\phi} \equiv B_r/B_\theta$ and $dr/d\phi = F_{\theta \phi} / F_{r
\theta} \equiv B_r/B_\phi$.

\begin{figure}
\centering
\includegraphics[height=.46\textwidth]{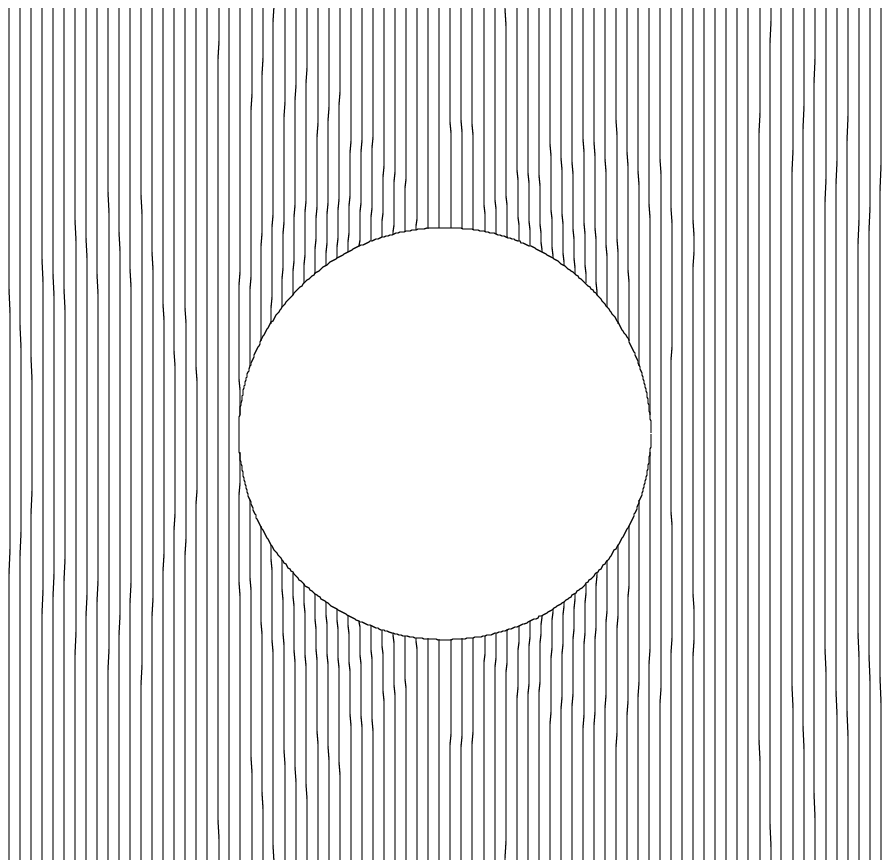}~~~\hfil
\includegraphics[height=.46\textwidth]{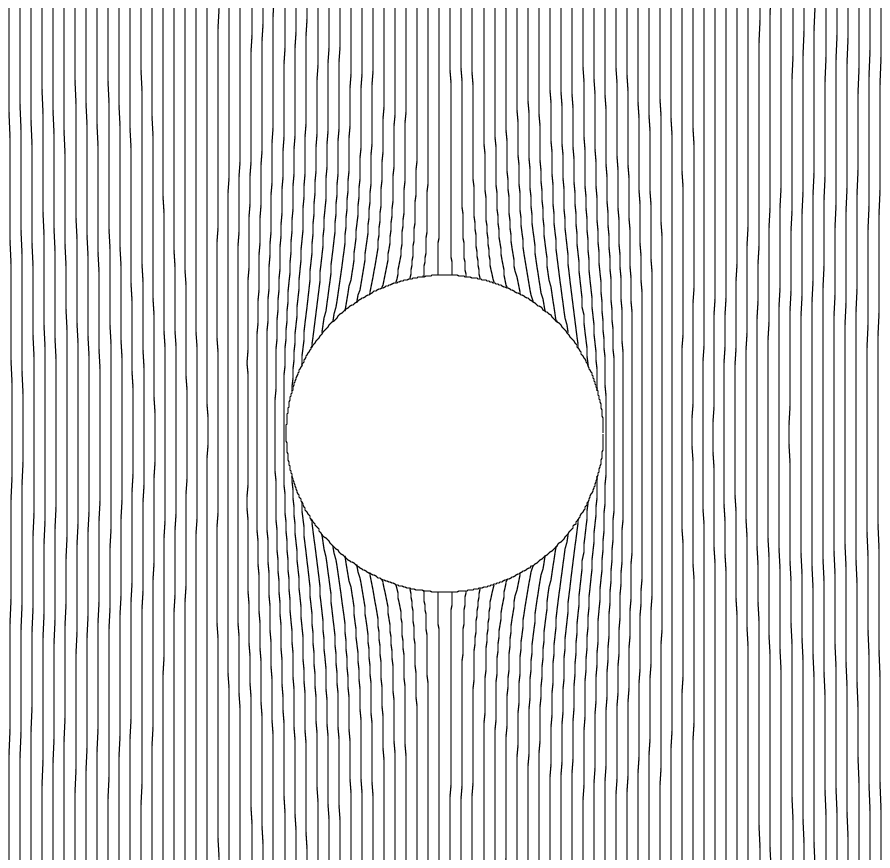}\\
(a) \hskip 6cm (b)\\
\includegraphics[height=.46\textwidth]{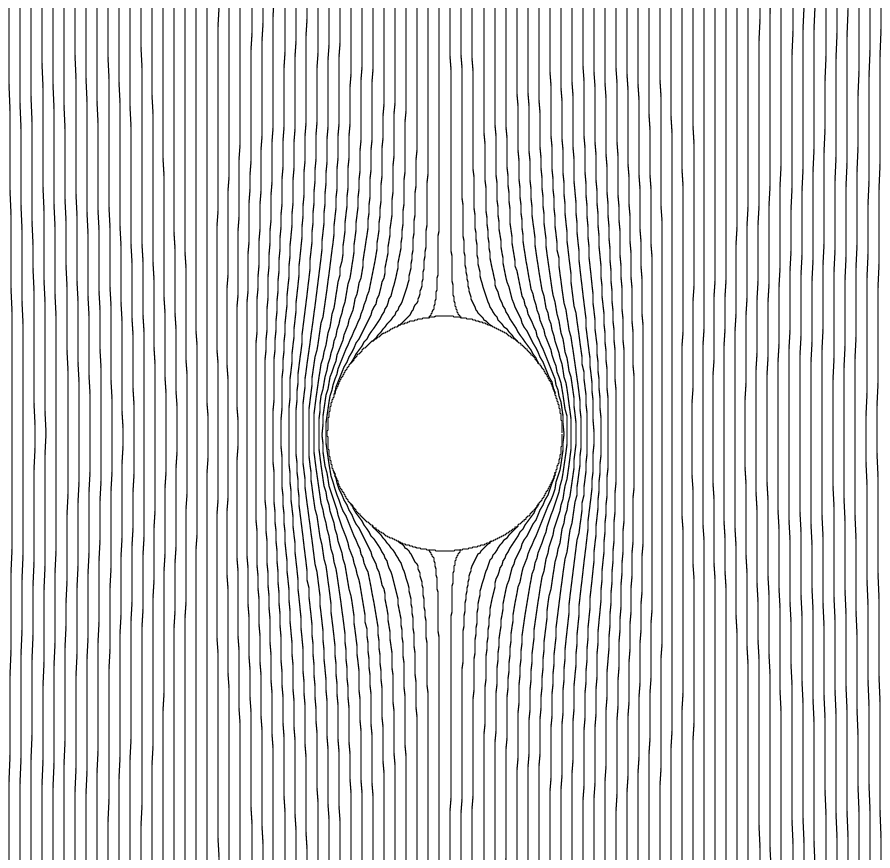}~~~\hfil
\includegraphics[height=.46\textwidth]{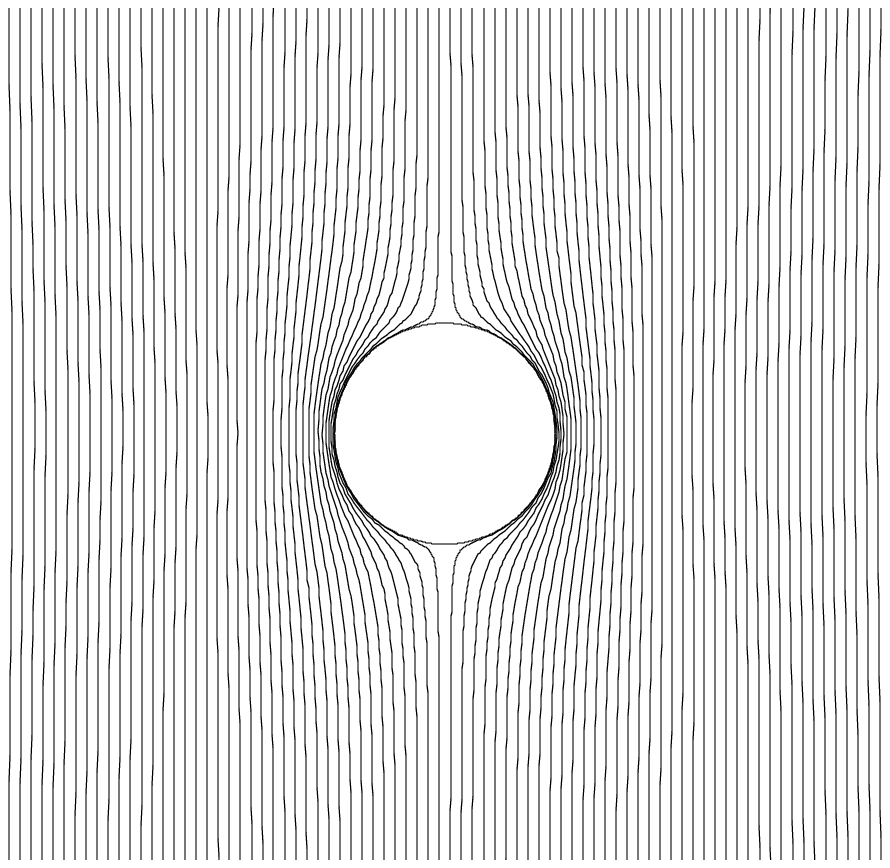}\\
(c) \hskip 6cm (d)
\caption{
Field lines of the test magnetic field homogeneous at infinity and aligned with hole's rotation axis.
Four cases with (a) $a=0.5 M$ (b) $a=0.9 M$ (c) $a=0.998 M$ and (d) $a=M$ are shown. Notice that  only a weak
flux expulsion takes place for the cases (a) and (b).
(Individual field lines start at $z=r \cos\theta = 4$; they are separated by $0.1 M$.)}
\label{Fig3}
\end{figure}

In the case of the aligned field we can easily verify (by using
$F_{\mu \nu}$ from \cite{BJ}) that the field lines lie on the
surfaces of constant flux,

\begin{equation}
\phi = \pi B_0 [\Delta + 2Mr \Sigma^{-1} \left( r^2 - a^2 \right) ]
\sin ^2 \theta = {\rm constant},
\label{9}
\end{equation}
where

\begin{equation}
\Delta = r^2 - 2Mr + a^2, \,\,\, \Sigma = r^2 + a^2 \cos ^2 \theta;
\label{10}
\end{equation}
$r, \theta, \phi$ are Boyer-Lindquist coordinates. The field
lines constructed numerically are shown in Figure \ref{Fig3}. The figures
clearly illustrate how the magnetic field is expelled from the horizon
when the angular momentum of the hole increases. Analogously to
the Reissner-Nordstr\"om case, no field line of the
asymptotically uniform magnetic field enters the horizon of an
extreme Kerr black hole.

\begin{figure}[b]
\centering
\includegraphics[width=.59\textwidth]{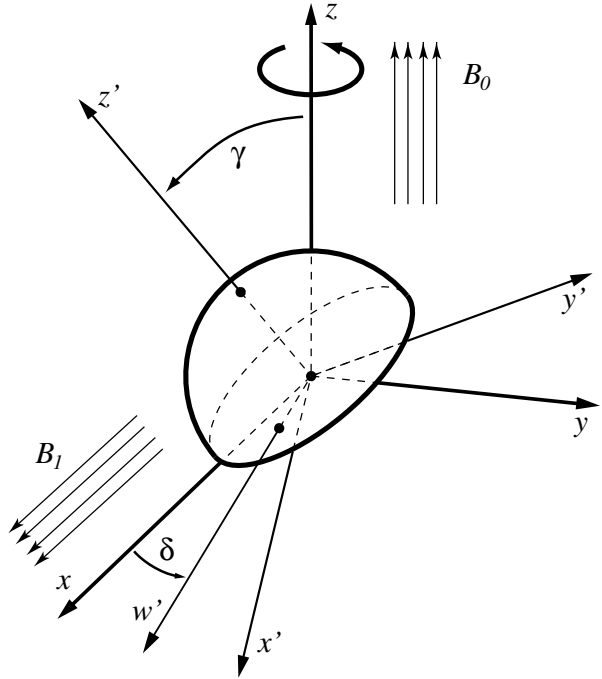}
\label{Fig4}
\caption{
A generally located hemisphere across which the magnetic flux of an asympotically uniform magnetic field is calculated.
The location of the hemisphere is specified by the angle $\gamma$ between the $z'$-axis
around which the hemisphere is rotationally symmetric and the $z$-axis around which the black hole rotates,
and by the angle $\delta$ between the projection of the $z'$-axis onto the equatorial plane and the $x$-axis
along which is aligned the component of field, $B_1$,
which is perpendicular to the rotation axis of the hole. $B_0$ is the component along this axis. (From \cite{BJ}.)}
\end{figure}

In the case of {\it general} stationary axisymmetric (i.e. ``aligned")
electromagnetic fields, one arives, using solutions given in \cite{BiD}
(in the Newman-Penrose formalism), at a similar result:
the flux of an arbitrary, axially symmetric stationary magnetic field across any part of the
horizon of an extreme Kerr black hole vanishes.

The structure of an asymptotically non-aligned field is much more
complicated. In this case $B_\phi \not= 0$ and the field lines are
dragged around the black hole. The lines, originally parallel to
each other, are twisted, some of them threading the horizon even
in an extreme case. The results of the numerical construction of
the magnetic lines asymptotically perpendicular to the rotation
axis of the Kerr black hole with $a/M=0.998$, as they look in the
equatorial plane when viewed from above, are given in \cite{KA}. The
field lines in Boyer-Lindquist coordinates are wound up around
the horizon; in the Kerr ingoing coordinates the field lines do not wind up.

The structure of the magnetic field near a black hole can be
characterized by the magnetic flux across (half of) the horizon.
A general position of the hemisphere can be specified on the
basis of an Euclidean picture. The magnetic flux across the
hemisphere can be defined invariantly as an integral over the
surface of the horizon \cite{BJ}

\begin{equation}
\Phi = \int^{2\pi}_{0} \!\!\!\!\! d \phi \int^{\Theta(\phi;\gamma ,
\delta)}_{0} \!\!\!\!\!\!\!\!\!\!\!\!d\theta~{F_{\theta \phi}}_{|r=r_+},
\label{14}
\end{equation}
where
$\Theta (\phi ; \gamma , \delta) \equiv \frac{\pi}{2} + \arctan
[\tan \gamma \cos (\phi-\delta)] $ - see Figure \ref{Fig4}.

Substituting for $F_{\theta \phi}$ and performing the integral we find rather complicated expression which,
however, can be understood intuitively in special cases: (i) If $\gamma=0$ the magnetic flux reduces to
$ \Phi = B_0  \pi r_+^2 (1-a^4/r_+^4)$; with $\gamma=0$ the total flux of the $B_1$ component of course vanishes.
As the hole becomes extreme we find $\Phi \rightarrow 0$ -- see Figure \ref{Fig3} (c-d).
(ii) If $\gamma = \pi/2$ the flux is $ \Phi = B_1 \pi[r_+^2 \cos \delta - (r_+ + M)a\sin \delta]$.
If the hole is not rotating one gets $ \Phi = B_1 \pi r_+^2 \cos \delta = \Phi_{\rm classical}$.
We get zero flux for $\delta=\pi/2$ and maximal one for $\delta = 0$. When the hole rotates the field lines are dragged along
and the flux vanishes across the hemisphere rotated by an angle $(\pi/2)-\delta_0$ where $\delta_0 \sim 27^o$ for $a=M>0$.
For a given $a$ there exists $\delta_{\rm max}$ for which the flux is maximal. With $a=M$ we find $\delta_{\rm max} \sim -63^o$,
$\Phi \sim 2.25 B_1 \pi r_+^2 = 2.25 \Phi_{\rm classical}$.
Hence, we see that although the flux of the aligned component $B_0$ decreases to zero with $a \rightarrow M$, the flux of the
perpendicular component is enhanced.

\section{Magnetic flux across the stretched horizon of an almost
extreme Kerr black hole}

In this section we shall outline some results of \cite{PR}
and of the discussions one of us (J.B.) had with Richard
Price in 1996. The discussions were concerned with the validity of
the membrane paradigm \cite{Th} for ``alex'' black holes, as we
call ``almost extreme'' black holes. It is well-known that a proper
distance from any point outside an extreme black hole to the outer
horizon is infinite. This fact implied the statement in
\cite{Th}, p. 120, that ``because the horizon is infinitely far
down in the [embedding] diagram, the finite magnetic flux has an
infinite spatial distance in which to wrap itself around the
embedding cylinder, and, consequently, near the horizon the
magnetic field falls off to zero.''

Notice, first, that in fact in a freely falling frame the
component $B_{(\theta)}$ and $B_{(\varphi)} = -B_{(\theta)}\cos \theta$ do
not vanish at the horizon of an extreme Kerr black hole, only the
radial component $B_{(r)}= 0$. (In case of an extreme
Reissner-Nordstr\"{o}m black hole all components of magnetic
field do vanish even in a freely falling frame at the horizon.)
Now in the spirit of the ``membrane paradigm'' consider a flux
across a stretched horizon characterized (in the Boyer-Lindquist
coordinates) by the redshift factor for locally non-rotating
frames (``lapse function''),

\begin{equation}
\alpha = \left[ (r^2 + a^2 \cos ^2 \theta) \Delta \right]
^{\frac{1}{2}} \left[ ( r^2 + a^2)^2 - a^2 \Delta \sin^2 \theta
\right] ^{-\frac{1}{2}},
\label{4}
\end{equation}
which at a stretched horizon is small, $\alpha_{Hs} = {\rm constant}
\ll 1$, but nonvanishing - in contrast to the true horizon for
which $\alpha_{Ht} = 0$. Introduce a parameter $\epsilon =
1-a/M$ so that $\epsilon$ is small for alex black holes and
vanishes for the extreme holes. We know that the magnetic flux of
axisymmetric fields vanishes at the true horizon in the limit $\epsilon
\to 0$:
\begin{equation}
\lim_{\epsilon \to 0} \phi_{|\alpha_{Ht}=0} = 0.
\end{equation}
A question arises whether if we ``exchange'' the limits, i.e.,
calculate first the flux across the stretched horizon of an alex
black hole and then go to the extreme case we could obtain a
nonvanishing quantity: will then

\begin{equation}
\lim_{\alpha_{Hs} \rightarrow 0} \left(~
   \lim_{\epsilon \rightarrow 0}
        {\phi_{|\alpha_{H_s\not=0}}} ~\right) \not= 0~?
\label{5}
\end{equation}
Restricting ourselves to asymptotically uniform fields we can
easily calculate the flux across the stretched horizon (using the
expressions for electromagnetic field given in \cite{BJ}) and
find that

\begin{equation}
\Phi \approx 8 \pi B_0 M^2 \alpha_H ~~ {\rm for} ~~
\epsilon \ll \alpha^2_H~,~~~~
\Phi \approx 4\pi B_0 M^2 (2\epsilon)^{\frac{1}{2}} ~~{\rm for}~~ \epsilon
\gg \alpha^2_H\,.
\label{6}
\end{equation}
Therefore, $\Phi$ across the stretched horizon is nonvanishing
even in the extreme case $\epsilon = 0$ but it depends on where
$H_s$ is located; however, $\Phi \rightarrow 0$ as $\alpha _{Hs}
\rightarrow 0$, so that the limit (\ref{5}) vanishes. For any $\epsilon
>0$ there exists $\delta > 0$ such that $\alpha_H<\delta$ and
$\Phi < \epsilon$. This conclusion seems to suggest that power in
the Blandford-Znajek model arises from regions with ``relatively
large'' $\alpha _H$ in near-extreme case.

\section{Black holes in magnetic fields: exact models}
We now turn to the exact stationary solutions of the
Einstein-Maxwell equations representing rotating, charged black
holes immersed in an axisymmetric magnetic field. In the
weak-field limit -- when $\beta \equiv B_0 M \ll 1$, the constant
$B_0$ in the weak-field limit characterizes the magnetic field
strength, $M$ the hole's mass -- there exists a region
$2M\ll r\ll {B_0}^{-1}$ where the spacetimes are approximately flat and
the magnetic field is approximately uniform. At $r\gg B_0^{-1}$ the
metrics approach Melvin's magnetic universe.

The simplest example is the magnetized Schwarzschild black hole:

\begin{equation}
ds^2 = \Lambda ^2 \{(1-2M/r)^{-1} dr^2 + r^2 d \theta ^2 -
(1-2M/r)dt^2\} + \Lambda ^{-2} r^2 \sin ^2 \theta \,\, d \phi^2,
\label{11}
\end{equation}
where $\Lambda = 1 + \frac{1}{4} B^2_0 r^2 \sin ^2 \theta$, the
electromagnetic field being given by $F_{\theta \phi} \equiv B_r = B_0
\Lambda ^{-2} r^2 \sin \theta \cos \theta$, $F_{r \phi} \equiv
-B_\theta = B_0 \Lambda^{-2} r \sin ^2 \theta$. The magnetic flux
across a general hemisphere on the horizon (with $\delta = 0$ due
to axisymmetry, cf. Figure \ref{Fig4}) reads as follows:

\begin{equation}
\Phi = 4 \pi \beta M \left[ 1+\beta ^2 \right] ^{-\frac{1}{2}}
\left[ 1+\beta ^2 + \tan ^2 \gamma \right] ^{-\frac{1}{2}}.
\label{12}
\end{equation}
For the flux across the ``upper half'' of the horizon $(\gamma =
0)$, we obtain the result given in \cite{BJ}. (See Eq. (41) therein where
incorrect $M$ should be replaced by $r_+ = 2M$ in the
denominator.)

If $\beta \ll  1$ the flux is equal to $4\pi \beta M (1+\tan ^2
\gamma)^{-\frac{1}{2}}$ in accordance with the weak-field
approximation. By increasing $\beta$ (with $M$ and $\gamma$ fixed)
$\Phi$ first increases, as we expect on classical grounds.
However, a further increase of $\beta$ leads to the decrease of
the flux. Indeed, there exists a value of the magnetic field
parameter $\beta$ for which the flux acquires its maximum value,
$\beta = (1+\tan ^2 \gamma)^{\frac{1}{4}}$. The global maximum, occurring
with $\gamma = 0$, is
\begin{equation}
\Phi=\Phi_{max}\equiv 2\pi M \doteq 3.22 \times 10^{39}
(M/10^9M _{\odot}) \left[ {\rm Gauss\;cm^2} \right].
\label{R13}
\end{equation}
The existence of the limiting value of the magnetic flux across
the horizon is caused by the gravitational effect of the magnetic field
which concentrates itself near the axis of symmetry when $\beta \gg 1$.

In \cite{K}, a rather exotic but interesting model was presented recently
in which a supermassive black hole is surrounded by a superstrong magnetic field
of the strength corresponding to just the half of the maximal flux (\ref{R13}).
In this model an electric induction field can accelerate charged particles up to an energy
$10^{18}$ GeV.

There exists a fairly extensive literature on exact solutions representing
rotating charged black holes in an external magnetic field (a magnetic Melvin-type universe).
They can be used to study the Meissner effect within the exact framework. One can make sure
that the magnetic fluxes vanish across the horizons of extreme black holes,
i.e. those with zero surface gravity. We refer to \cite{BiKa} and to very recent paper \cite{KaBu} for the review
and relevant references.

\section{Meissner effect for superconducting branes and extremal black
holes in string theory}

In 1998 a comprehensive paper by Chamblin, Emparan and Gibbons
\cite{CEG} reviewed and developed evidence for the Meissner
effect for extremal black hole solutions in string theory and
Kaluza-Klein theory. Here we shall mention some of their ideas
and results, sticking closely to their discussion. They first
give a brief phenomenological description of the Meissner effect
in superconductors. Using London equation
\begin{eqnarray}
\vec{j} = - \,{\rm constant}~ \vec{A}~,
\label{7}
\end{eqnarray}
($\vec{j}$ - current, $~\vec{A}$ - vector potential),
and Maxwell's equations, one arrives at the equation
\begin{equation}
\nabla ^2 \vec{A} = \lambda ^2 \vec{A},
\label{8}
\end{equation}
$\lambda = {\rm constant}$, given by constants characterizing the
material. The solution in one dimension, $A \sim \exp(-\lambda
x)$, indicates that $\vec{A}$ (and thus the magnetic
field as well) decreases exponentially as one goes from the
surface into the superconductive material. This is the classical
Meissner effect. We see that the expulsion of a magnetic field
from extreme (standard) black holes is analogous -- but there is a
difference: no magnetic flux at all penetrates into a hole. The
relativistic generalization of the London equation reads

\begin{equation}
J_{\mu} = - \lambda^{-2} A_{\mu} + \partial _{\mu}
\Lambda,
\label{RLON}
\end{equation}
or

\begin{equation}
-\lambda ^{-2} F_{\mu \nu} = \partial _{\mu} J_{\nu} - \partial
_{\nu} J_{\mu},
\label{RLON2}
\end{equation}
where $J_\mu$ is $4$-current and $A_\mu$ 4-potential.
In \cite{CEG} equation (\ref{RLON2}) is adopted as the criterion
for superconductivity. The authors first review the work of
Nielsen and others showing that this equation is satisfied in the
Kaluza-Klein theory on the world volume of extended objects
carrying Kaluza-Klein currents. Then they consider self-gravitating
extended superconducting objects and demonstrate how the flux of
a gauge field is expelled from the intersection of two sets of
6-dimensional self-gravitating branes of 11-dimensional supergravity.

Several examples are then constructed in \cite{CEG} demonstrating
that the expulsion of gauge fields occurs always at the extreme
horizon. Consider a string in $d=5$ which is wrapped along the
string direction so that a dilatonic black hole solution in $d=4$
arises. The starting metric in $d=5$ reads

\begin{equation}
ds^2 = H^{-\frac{1}{3}} \left( -f dt^2 + dz^2 \right) +
H^{\frac{2}{3}} \left[ f^{-1} dr^2 + r^2 \left( d\theta ^2 + \sin
^2 \theta d \varphi ^2 \right) \right],
\label{15}
\end{equation}
where

\begin{equation}
H = 1+q/r, \,\,\, f=1-r_0/r,
\label{16}
\end{equation}
$q, r_0$ are constants, an event horizon being at $r=r_0$;~ $r_0=0$
is the extreme case. If this geometry is compactified along the
string direction $z$ in such a way that the compactification
direction is twisted -- the compactification is made along the orbits of the vector
$(\partial/\partial z) + B(\partial/\partial\varphi)$, where
constant $B$ will describe the asymptotic value of the magnetic
field along the axis -- one obtains a black hole and the
Kaluza-Klein magnetic field. The field is described by the
potential whose $\varphi$ - component is given by

\begin{equation}
{\cal {A}} = B g_{\varphi \varphi} / (g_{zz} + B^2 g_{\varphi
\varphi}) = B Hr^2 \sin ^2 \theta / (1+B^2 Hr^2 \sin ^2 \theta).
\label{17}
\end{equation}
In general one finds a non-vanishing flux across the horizon at
$r=r_0$. In the limit of an extreme hole, however, at the horizon
$r=0$ the potential ${\cal{A}}$ vanishes and no magnetic flux
thus penetrates the horizon.

As another example of the Meissner effect Chamblin, Emparan and
Gibbons consider the field expulsion from extreme rotating black
holes. First they recall the expulsion of the asymptotic uniform
magnetic test field in the extreme Kerr black hole background
discussed in Section 3. Then they construct an exact solution (by
taking the product of the standard Kerr solution with
$x^5$-direction and applying a twisted reduction procedure,
similar to that considered above but involving also a twist in
the time coordinate) representing a Kerr black hole in the {\it
exact} Kaluza-Klein gauge field. Again, this exact field exhibits
the Meissner effect in the extreme case.

At the end of the subsection (see \cite{CEG}, {\it V.} A.) the authors
write that they considered the solutions in which the magnetic
field is aligned with the rotation axis of the black hole but
that according to our work \cite{BD} the Meissner expulsion can
also be seen for fields where no alignment is assumed. This is
not correct: as mentioned in the previous Section 3, and
demonstrated in detail in \cite{BJ}, the asymptotically
non-aligned fields do penetrate the extreme Kerr horizons. Only
general {\it axisymmetric, stationary} fields on the Kerr
background exhibit the Meissner effect in the extreme limit. On
the other hand, it is well known that configurations with
stationary external fields which are not axisymmetric, are not
stable -- due to the torque exerted on the horizon by an external
non-axisymmetric field (see e.g. \cite{Th}) -- and evolve towards
axisymmetric configurations.

In case of the standard superconductivity, when the Meissner
effect arises, the field inside a superconductor becomes a pure
gauge. This is not the case with extreme black holes. In
\cite{BD} in Figure 1(d) the field lines are constructed inside
the extreme Reissner-Nordstr\"om horizon. Clearly, one cannot
claim that the interior of extreme black holes is in a
superconducting state.

The question of the flux expulsion from the horizons of extreme
black holes in more general frameworks is not yet understood
properly. The authors of \cite{CEG} ``believe this to be a
generic phenomenon for black holes in theories with more
complicated field content, although a precise specification of
the dynamical situations where this effect is present seems to be
out of reach.'' That for abelian Higgs vortices the phenomenon of
flux expulsion from extreme black holes does not occur in all
cases has been argued analytically and investigated numerically
recently \cite{BEG}. In particular, it appears that thin
cosmic strings (modelled by the vortices) can pierce the extreme
horizons whereas a thicker string will be expelled.

\vskip 3mm
{\bf Acknowledgments}. The authors are grateful to the organizers of the 3rd ICRA
Workshop for their hospitality.
Support from the grant No. GA\v CR 202/99/0261 of the Czech Republic and
the grant GAUK 141/2000 of the Charles University is also gratefully acknowledged.

\end{document}